\documentclass[prc,superscriptaddress,showpacs,twocolumn]{revtex4-1} 
\usepackage{amssymb}
\usepackage{amsmath}
\usepackage{color}
\usepackage{graphicx}
\usepackage{soul}
\usepackage{etoolbox}
\patchcmd{\section}
  {\centering}
  {\raggedright}
  {}
  {}
\patchcmd{\subsection}
  {\centering}
  {\raggedright}
  {}
  {}

\newcommand{\muorb}{\ensuremath{\mu_{\mathrm{orb}}}}
\newcommand{\DSO}{{\ensuremath{\Delta_{\text{SO}}}}}
\newcommand{\DKK}{{\ensuremath{\Delta_{\text{KK'}}}}}

\newcommand{\SUF}{\ensuremath{SU(4)}}
\newcommand{\SUT}{\ensuremath{SU(2)}}

\newcommand{\dop}{\hat{d}}

\newcommand{\Hop}{\hat{H}}

\newcommand{\KK}{\text{KK'}}
\newcommand{\SO}{\text{SO}}
\newcommand{\dtxt}{\text{d}}
\newcommand{\CNT}{\text{CNT}}
\newcommand{\B}{\text{B}}

\begin{document}

\title{Blocking transport resonances via Kondo many-body entanglement in quantum dots}
\author{Michael Niklas}
\affiliation{Institute for Theoretical Physics, University of
Regensburg, 93040 Regensburg, Germany}

\author{Sergey Smirnov}
\affiliation{Institute for Theoretical Physics, University of
Regensburg, 93040 Regensburg, Germany}

\author{Davide Mantelli}
\affiliation{Institute for Theoretical Physics, University of
Regensburg, 93040 Regensburg, Germany}

\author{Magdalena Marga\'nska}
\affiliation{Institute for Theoretical Physics, University of
Regensburg, 93040 Regensburg, Germany}

\author{Ngoc-Viet Nguyen }
\affiliation{ Institut N\'{e}el, CNRS and Universit\'{e} Grenoble Alpes, 
38042 Grenoble, France}

\author{{Wolfgang Wernsdorfer}}
\affiliation{ Institut N\'{e}el, CNRS and Universit\'{e} Grenoble Alpes,
38042 Grenoble, France}

\author{Jean-Pierre Cleuziou}
\affiliation{ Institut N\'{e}el, CNRS and Universit\'{e} Grenoble Alpes, 
38042 Grenoble, France}

\affiliation{CEA and Universit\'{e} Grenoble Alpes, INAC-SPSMS,
38054 Grenoble, France}

\author{Milena Grifoni*}
\affiliation{Institute for Theoretical Physics, University of
Regensburg, 93040 Regensburg, Germany}

\date{\today}

\begin{abstract}
Many-body entanglement  is at the heart of the Kondo effect, which has its hallmark in quantum dots as a zero-bias conductance peak at low temperatures. It signals the emergence
of a conducting singlet state formed by a localized dot degree of freedom and conduction electrons.   
Carbon nanotubes offer the possibility to study the emergence of the Kondo entanglement  by tuning many-body correlations with a gate voltage. Here we  show an undiscovered side of 
 Kondo correlations,  which counterintuitively  tend to block conduction channels:   inelastic cotunneling lines in the magnetospectrum of a carbon nanotube strikingly disappear when tuning the gate voltage. Considering the global \SUT\ $\otimes $ \SUT\ symmetry of a  nanotube coupled to leads, we find that only resonances  involving flips of the Kramers pseudospins, associated to this symmetry, are observed at temperatures and voltages below the corresponding Kondo scale. Our results demonstrate the robust formation of  entangled many-body states with no net pseudospin.
\end{abstract}

\maketitle

The ubiquity of Kondo resonances in quantum dots relies on the fact that their occurrence  requires only  the presence of degenerate dot states, whose degeneracy is associated to  degrees of freedom which are conserved during the tunneling   onto and out of the dot \cite{Hewson1997}.  Finite magnetic fields can be used   to break  time-reversal symmetry related degeneracies and  
  unravel the deep nature of the Kondo state by tracking the magnetic field evolution of split Kondo peaks \cite{G-Gordon1998,Nygaard2000,Sasaki2004,nature-jarillo:484,Jarillo-Herrero2005,Quay2007,Makarovski2007,Grap2011,Lan2012,Tettamanzi2012}.
In a recent work \cite{Schmid2015}, the striking report was made that specific transport resonances were not observable in nonlinear magnetoconductance measurements of split Kondo peaks in carbon nanotubes (CNTs), despite being expected from theoretical predictions \cite{Choi2005,Fang2008,Fang_Erratum}. Even more intriguing is that those resonances were recorded in inelastic cotunneling measurements in the weak coupling regime \cite{Jespersen2011}.  
Because in \cite{Schmid2015} no comparative measurement for the weak-coupling regime was reported, the missing of resonances could not be unambigously interpreted as a signature of the Kondo effect.   
 From a closer inspection of other experimental reports for the Kondo regime \cite{nature-jarillo:484,Quay2007,Lan2012,Cleuziou2013}, we notice that the absence of some resonances seems systematic.

In the following we study the low-temperature nonlinear electron transport in a very clean CNT quantum dot \cite{Laird2015}. By simply sweeping a gate voltage \cite{nmat-cao:745,Makarovski2007}, we could tune the same CNT device from a weak coupling regime, where Coulomb diamonds and inelastic cotunneling are observed, to a Kondo regime with strong many-body correlations to the leads. Then, using nonlinear magnetospectroscopy, transport resonances have been measured.
The two regimes have been  described using accurate transport calculations based on perturbative  and nonperturbative approaches in the coupling, respectively.
The missing resonances  in the Kondo regime have been clearly identified, and their suppression fully taken into account by the transport theory. Accounting for 
both spin and orbital degrees of freedom, 
we discuss a global  \SUT\ $\otimes $ \SUT\ symmetry related to the presence of two Kramers pairs in realistic carbon nanotube devices with spin-orbit coupling (SOC) \cite{Ando2000,Kuemmeth2008,prb-delvalle:165427,ncomms-steele:1573} and valley mixing \cite{Kuemmeth2008,Jespersen2011,GroveRasmussen2012,Izumida2015,Marganska2015}.
In virtue of an effective exchange interaction,  virtual transtions which  flip the Kramers pseudospins  yield low-energy many-body singlet states with net zero Kramers pseudospin. 
This result in turn reveals that the transport resonances  suppressed in the deep Kondo regime are associated to virtual processes which do not flip the Kramers pseudospin.

\section*{\large R\MakeLowercase{esults}}
{\bf {Measurement and modelling of transport regimes.}} 
The device under study consists of a semiconducting CNT, grown \textit{in-situ} on top of two platinum contacts, used as normal metal source and drain leads. Details of the device fabrication were reported previously \cite{Cleuziou2013} (see also the Methods). The CNT junction is suspended over an electrostatic gate and can be modelled as a single semiconducting quantum dot of size imposed by the contact separation ($\approx$ 200 nm). All the measurements were performed at a mixing chamber temperature of about $T_{\rm exp}$= 30 mK,  which sets a lower bound to the actual electronic temperature.  The set-up includes the possibility 
 to fully rotate an in-plane magnetic field up to 1.5 T.

The CNT level spectrum is depicted in Figs. 1a and 1b. 
Transverse bands, represented by the coloured hyperbolae in Fig. 1a, emerge from the graphene Dirac cones as a consequence of 
 the quantization of the transverse momentum $k_\perp$. Bound states (bullets)  are due to the quantization of the longitudinal momentum $k_\parallel$. Four-fold spin-valley degeneracy   yields the exotic spin plus orbital \SUF\ Kondo effect \cite{Choi2005,nature-jarillo:484,Jarillo-Herrero2005,Makarovski2007,Anders2008,Cleuziou2013,Ferrier2015}.
The  SOC removes the spin degeneracy of  the transverse bands in the same valley (red and blue hyperbolae), and hence the \SUF\ symmetry \cite{Jarillo-Herrero2005,Fang2008,Fang_Erratum,Galpin2010,Lan2012,Cleuziou2013,Schmid2015,Ferrier2015}.    
Due to time reversal  symmetry, for each  $k_\parallel$  a quartet of states consisting of two Kramers pairs splitted by the energy $\Delta=$ \DSO\ arises. 
 When also valley mixing is present, with the energy scale \DKK,  orbital states are formed which are superpositions of valley states. A quartet now consists of  two  Kramers doublets at  energies $\varepsilon_{\rm d}=\pm\Delta/2$, with $\Delta=\sqrt{\DSO^2+\DKK^2}$, see Fig. 1b.

By sweeping the gate voltage, the chemical potential is moved from above (electron sector) to below (hole sector) the charge neutrality point and quadruplets of states are thus successively emptied.
 This pattern is visible in a typical measurement of the differential conductance $dI/dV$ versus the bias voltage $V_{\rm sd}$ and the gate voltage $V_{\rm g}$, Figs. 1c, 1d, which exhibits a characteristic four-fold periodicity. Figure 1c displays such a stability diagram for the electron sector, where Coulomb diamonds and inelastic cotunneling excitation lines are visible. Owing to significantly different  ratios $\Gamma/U$ of the tunnel coupling to the charging energy   in the valence and conduction regimes,  Kondo physics dominates for odd hole number in the hole sector shown in Fig. 1d. 

In order to investigate the dominant transport mechanisms, we have performed transport calculations for both regimes, using a standard minimal model for a longitudinal mode of a CNT 
quantum dot with SOC and valley mixing terms \cite{Jespersen2011,Laird2015}.
  The explicit form of the model Hamiltonian $\hat H_{\rm CNT}$  and the parameters used  for the transport calculations are provided in  the  Methods. The transport calculations in the electron regime implement  a perturbation theory (PT)  which  retains all tunneling contributions to the dynamics of the CNT reduced density matrix up to second order in the tunnel coupling $\Gamma$. This approximation thus accounts for Coulomb blockade (first order in $\Gamma) $ and leading order cotunneling processes (second order in $\Gamma$), and it is expected to give accurate results for small ratios $\Gamma/k_{\rm B}T$ and $\Gamma/U$ \cite{Koller2010}.
The results of the calculations for the differential conductance are shown in Fig. 1g; a gate trace in Fig. 1e. The perturbative theory reproduces the position of the inelastic cotunneling thresholds (panels 1c and 1g).  In the gate trace of Fig. 1e the experimental peaks are wider than the theoretical ones. Because in the latter the broadening is solely given by the temperature,  this indicates that higher order terms 
are responsible for a broadening of the order $\Gamma$  and for a Lamb shift  of the experimental peaks \cite{Koenig1996,Pedersen2005,Dirnaichner2015}. 
In this work we are interested only in the evolution of the cotunneling resonances in magnetic field, which is well captured by the  perturbative approach as long as Kondo ridges have not yet formed.

This situation radically changes in the hole sector where the gate trace reveals Kondo ridges for odd hole numbers. The theoretical trace in Fig. 1f is the outcome of a nonperturbative numerical DM-NRG calculation \cite{Bulla2008} which uses the same model Hamiltonian  but with slightly different parameters.
The strong suppression of the conductance in the valley with even hole occupancy is an indication of the breaking of  the \SUF\ symmetry in the presence of SOC and valley mixing to an $\SUT\otimes\SUT$ one \cite{Galpin2010,Mantelli2015}. In the DM-NRG  calculations the two-particles exchange $J$ was not included due to high computational costs. The latter further reduces the symmetry  in the 2h valley  (see e.g the spectrum in Fig. 2b), and hence the experimental conductance is more rapidly suppressed in that valley than as predicted by our simulations. On the other hand, $J$ is not relevant for describing the spectrum in the 3h and 1h cases (Figs. 2a, 2c), which is the focus of the present work. 

In the DM-NRG calculations the fit to the experiment was done assuming a temperature of $T=30$ mK.   From the so extracted parameters we evaluate the temperature dependence of the conductance 
at  $-\varepsilon_{\rm d} =U/2 -\Delta/2$,  and $-\varepsilon_{\rm d}=5U/2 +\Delta/2$,  corresponding to gate voltage values located roughly in the middle of the 1h and 3h valleys, respectively, and extract the Kondo temperatures, see Fig. 1h.  At such values of $\varepsilon_{\rm d}$ the Kondo temperature takes its minimal value in a given valley, which sets a lower bound for $T_{\rm K}$ \cite{Mantelli2015}. We find $T_{\rm K}=84$ mK and $T_{\rm K}=160$ mK for the 1h and 3h valleys, respectively.  Correspondingly,    $0.1 <T_{\rm exp}/T_{\rm K}<1$, suggesting that the experiment is in the so-called Kondo crossover regime  \cite{Hewson1997} also for the actual electronic and Kondo temperatures. \\

{\bf Virtual transitions revealed by magnetospectroscopy.}
Having set the relevant energy scales for both the electron and hole sectors, we proceed now with the investigation of 
 magnetotransport measurements at finite source-drain bias, which  have been performed for different fillings.  
A magnetic field ${\bf B}$  breaks time reversal symmetry and thus the Kramers degeneracies. By performing
inelastic cotunneling spectroscopy,  we can get information on the lowest lying resonances of our interacting system.
The magnetospectrum corresponding to  electron filling $n_{\rm e}=1,2,3$ of a longitudinal quadruplet, as expected for the perturbative regime,  is shown in Figs. 2a - 2c.  
For the case of odd occupancies, we call    ${\cal T}$ transitions processes within a Kramers pair;  $\cal{C}$ and $ \cal{P}$ operations are associated to  inter-Kramers  transitions, as shown in Figs. 2a and 2c.
Panels 2d-2f and 2g-2i show magnetotransport measurements and theoretical predictions for the electron and hole regimes, respectively.
In these panels the current second derivative d$^2I/{\rm d}V^2$ is reported. We have preferred this quantity over the more conventional d$I/{\rm d}V$ (shown in the Supplementary Figures 4 and 5 and discussed in the Supplementry Note 4) to enhance eye visibility of the excitation spectra. 
 In panels 2d-2f as well as 2h we have used our perturbative approach \cite{Koller2010}.
The calculations in Figs. 2g, 2i, in contrast, are based  on the Keldysh effective action (KEA) method \cite{Smirnov_2013,Smirnov_2013a} and are non-perturbative.
The nature of the dominant inelastic transitions is clearly identified by simply looking at the excitation spectrum (dashed lines in Figs. 2d-2i). All inelastic transitions from the ground state are resolved in the cotunneling spectroscopy performed in the low coupling electron regime, similar to previous reports \cite{Jespersen2011}. 
When inspecting the hole regime, though, it is clear that only for the 2h case, panel 2h, the experimental data can be interpreted by means of a simple cotunneling excitation spectrum; moreover,  the 2e and 2h  cotunneling spectra are very similar.
In the 1h and 3h cases shown in panels 2g, 2i   Kondo correlations dominate the low energy transport, and differences with respect to the electron sector are seen. 
The zero-bias Kondo peak does not immediately split as the field is applied; rather the splitting occurs at a critical field such that the energy associated to the inelastic $\cal{T}$ transition is of the order of the Kondo temperature \cite{Hewson1997}.  In the 1h valley the lowest pair of levels merges again for values of the field of about 1.2 Tesla, yielding a Kondo revival \cite{nature-jarillo:484,Galpin2010}. Bias traces of the differential conductance highlighting the revival are shown in the Supplementary Figure 3 and analyzed in the Supplementary Note 3. 
Striking here is the observation that, in contrast to the 1e and 3e cases, {\em only one} of the two inter-Kramers transitions is resolved in the experimental data for the 3h and 1h valley. However,  in particular for the 1h case,  the  $\cal{P}$ and ${\cal C}$ excitation lines as expected from the excitation spectrum should be separated enough to be experimentally distinguishable, similar to the 3e case. By comparing with the excitation spectrum (dashed lines in panels 2g, 2i), we conclude that it is the ${\cal P}$ transition which is not resolved.  Our KEA transport theory qualitatively reproduces these experimental features.  \\
Magnetotransport measurements performed for other quadruplets both in the conduction and valence regimes exhibit qualitatively similar features (see Supplementary Figures 6-8, Supplementary Table 1 and  Supplementary Note 5), and hence confirm the robustness of the suppression of  $\cal{  P}$ transitions in the Kondo regime.
 Our results naturally reconcile the apparently contradictory observations in Refs. \cite{Schmid2015} and \cite{Jespersen2011}. Furthermore, they suggest that the inhibition of selected resonances
in the Kondo regime is of fundamental nature. 

{\bf Fundamental symmetries of correlated CNTs.}
To understand the experimental observations microscopically, we have analyzed those symmetries of an isolated CNT which also hold in the presence of on-site Coulomb repulsion typical of Anderson models. \\
 \indent In the absence of a magnetic field, one finds a  $U(1)\otimes U(1)\otimes SU (2) \otimes SU (2)$ symmetry related  to the existence of two pairs of  time-reversal degenerate doublets,  see Fig. 1b, called in the following upper $({\rm u})$ and lower $({\rm d})$  Kramers channels.
The $U(1)$ symmetries reflect charge conservation in each Kramers pair with generators 
$\hat Q_\kappa=\frac{1}{2}\sum_{j \in \kappa}(\hat n_j-\frac{1}{2})$ which 
 measure the charge of the pair with respect to the half-filling. Here is $j= (1,2)$ or $ (3,4)$ for $\kappa={\rm u}$ or ${\rm  d}$. The \SUT\ symmetries are generated by the  spin-like operators 
${ \hat{\bf J}_\kappa}=\frac{1}{2}\sum_{j,j'\in \kappa}\hat d^\dagger_j {\boldsymbol\sigma}_{j,j'}\hat d_{j'}$. Here ${\boldsymbol\sigma}$ is the vector of Pauli matrices. 
 Physically, $\hat J^z_{\rm u}=(\hat n_1-\hat n_2)/2$ and   $ \hat J^z_{\rm d}=(\hat n_4-\hat n_3)/2$ account for the charge unbalance within the Kramers pair.  Thus, an isolated CNT with one electron or a hole only in the quadruplet has a net Kramers pseudospin (and charge).   Fig. 3a shows the { two} degenerate groundstate configurations  $\vert\Downarrow;-\rangle_{}$, $\vert\Uparrow;- \rangle_{}$  of the isolated CNT with an unpaired  effective spin ($\Downarrow$ or $\Uparrow$)  in the lowest Kramers pair and no occupation (symbol "-") of the upper Kramers pair.
In the weak coupling regime, a perturbative approach to linear transport accounts for elastic cotunneling processes involving the doubly degenerate groundstate pair \cite{Grabert}. These  virtual transitions are denoted ${\cal I}$ or ${\cal T}$ when they involve the same state or its Kramers partner, respectively,  see Fig. 3a.  
A finite magnetic field breaks the \SUT\ symmetries. However, former degenerate CNT states can still be characterized according to the eigenvalues of the $\hat Q_\kappa$ and $\hat J^z_\kappa$ operators, since they commute with the single-particle CNT Hamiltonian which has in the Kramers basis the form (see Methods):
\begin{equation}
\hat H_{0}=\sum_{\kappa =\pm}\left(\bar\varepsilon ({\bf B})+\kappa \frac{\bar\Delta ({\bf B})}{2}\right) \hat N_\kappa + (2\delta \varepsilon ({\bf B})+\kappa\delta \Delta ({\bf B})) \hat J^z_\kappa ,
\end{equation}
where u/d$=+/-$, $\hat N_\kappa=2\hat Q_\kappa +1$, and at zero field is $\bar\Delta({ B}=0)=\Delta$, $\bar\varepsilon(B=0)=\varepsilon_{\rm d}$, $\delta\varepsilon=\delta\Delta=0$.
Hence our finite-bias and finite magnetic field spectroscopy allows us to clearly identify the relevant elastic and inelastic virtual processes according to the involved Kramers charge and spin.
As illustrated in Fig. 3b, in the weak tunneling regime only energy differences matter, and hence both intra-Kramers (${\cal I}$, ${\cal T}$) and inter-Kramers (${\cal P}$, $\cal{C}$)   transitions are expected in transport. 
In the Kondo regime this picture  changes. As we shall demonstrate, emerging Kondo correlations lead to the  progressive screening of
the Kramers  pseudospin of the dot 
by the conduction electrons. 

To this aim we observe that, when a sizeable tunnel coupling to the leads is included, 
the CNT charge and pseudospin operators $\hat Q_\kappa$ and $\hat {\bf J}_\kappa$ are no longer symmetries of the coupled system, since the tunneling does not conserve the dot particle number. 
The occurrence of the Kondo effect, however, suggests that the  CNT quantum numbers $j=1,2,3,4$  are carried also by the conduction electrons and conserved during tunneling \cite{Choi2005}. This is the case when the dot is only a segment of the CNT  (see Supplementary Figure 1). Following \cite{Mantelli2015}, we hence introduce charge, $\hat{\cal { Q}}_\kappa=\hat Q_\kappa +\hat Q_{{\rm L},\kappa}$, and pseudospin, $ \hat  {\boldsymbol{\cal  J}}_{\kappa}$=$\hat  {{\bf  J}}_{\kappa}+\hat  {{\bf  J}}_{{\rm L},\kappa}$, operators of the coupled  CNT plus  leads  (L) system. 
Under the assumption that the tunneling couplings are the same within each Kramers channel $\kappa = \rm{u,d}$, the total Hamiltonian (see Supplementary Methods) commutes with the charge and pseudospin operators  $\hat{\cal { Q}}_\kappa$ and $ \hat  {\boldsymbol{\cal  J}}_{\kappa}$, which hence  generate a $U(1)\otimes U(1)\otimes SU(2) \otimes SU(2)$ symmetry of the coupled system.   As a consequence,  many-body states can be characterized by the quadruplet  of eigenvalues $({\cal Q}_{\rm d},{\cal Q}_{\rm u}; {\cal J}_{\rm d},{\cal J}_{\rm u})$, where the highest eigenvalue ${\cal J}_\kappa$  of $\hat {\cal J}^z_\kappa$ is indicated in the quadruplet. This notation gives direct access to the eigenvalues $    {\cal  J}_{\kappa}  ( {\cal  J}_{\kappa}+1) $ of $ \hat  {\boldsymbol{\cal  J}}_{\kappa}^2$. Such quadruplets can be numerically calculated within our 
scheme for the Budapest DM-NRG code \cite{Thot2008}, and yield (for the valleys with one electron or one hole)  a \textit{singlet} ground state characterized by the quadruplet $(0,0;0,0)$. Thus "0" is also eigenvalue of ${\hat {\boldsymbol{\cal J}}}_{\rm u}^2$ and ${\hat{\boldsymbol{\cal J}}}_{\rm d}^2$.  
 {\em I.e., we find  a unique ground state with no net pseudospin.} 
This situation is illustrated in Fig. 3c: due to ${\cal Q}_\kappa=0$, the Kramers channels are half-filled (two charges per channel), whereby one charge arises from the electron trapped in the CNT itself. For $\Delta=0$ this CNT charge   is equally distributed among the two channels, while for large values of $\Delta/T_{\rm K}(\Delta)$, as in our calculation (see Fig. 1h), it is  mainly in the lowest Kramers channel. Thus at zero temperature the localized CNT pseudospin is fully screened by an  opposite net pseudospin in the leads. 
In the orthonormal basis $\{ \vert m\rangle_{}\otimes\vert n\rangle_{\rm L}\}$  spanned by the pseudospin eigenstates of CNT and leads this ground state is characterized by   the entangled configuration $\frac{1}{\sqrt{2}}[\vert\Uparrow;-\rangle_{}\otimes\vert\Downarrow;\Downarrow,\Uparrow\rangle _{\rm L}-\vert\Downarrow;-\rangle_{}\otimes\vert \Uparrow;\Downarrow,\Uparrow \rangle_{\rm L}]$ of dot and leads  pseudospins. 

In the standard spin-1/2 Kondo effect the appearance of a unique singlet ground state with no net spin is  the result of the screening of  the quantum impurity spin by the conduction electrons spins,  due to the antiferromagnetic character of the coupling constant between such degrees of freedom \cite{Hewson1997}. Triplets are (highly) excited states of the system.
To interpret the spin 1/2 Kondo effect in quantum dots, it is possible to derive from an Anderson model an effective Kondo Hamiltonian \cite{Schrieffer1966} given by the product of the quantum dot spin and the conduction electrons spin. The coupling constant for this product is positive and thus antiferromagnetic. Also for the more complex case of a CNT effective Kondo Hamiltonians have been derived, with positive coupling constants for Kramers channels identified by orbital and spin degrees of freedom \cite{Choi2005,Lim2006}. The antiferromagnetic character of the coupling constants remains also when, as in our case, the more abstract Kramers pseudospin is used.

A natural consequence of the antiferromagnetic nature of the correlations is that at low temperatures and zero bias elastic virtual transitions which flip the pseudospin, i.e., ${\cal T}$ transitions, are  favoured, as depicted in Fig. 3c. Similarly,  ${\cal C}$ transitions are inelastic processes which flip the pseudospin and become accessible at finite bias, as shown in Fig. 3d.  They connect the singlet ground state to an  excited state where the CNT charge is located in the upper Kramers channel. Our results suggest that $ {\cal P}$ transitions are inhibited  because they involve virtual transitions which conserve the pseudospin.\\

{\bf Entanglement of Kramers pseudospins.}
To further confirm that it is the Kramers pseudospins  and not  distinct spin or orbital degrees of freedom which should be considered in the most general situations, we report results for the differential conductance as a function of the angle $\theta$ formed by the magnetic field and the CNT's axis. 
The combined action of SOC, valley mixing and non collinear magnetic field mixes spin and valley degrees of freedom which, in general, are no longer good quantum numbers to classify CNTs states. Nevertheless, the three discrete ${\cal T}$, $\cal{ P}$ and $\cal{C}$ operations still enable us to identify the inelastic transitions in the 1h and 3h case, {\em independent} of the direction of the magnetic field.
 The angular dependence of both energy and excitation spectrum for a fixed magnetic field amplitude is shown in Figs. 4a, 4b for the 3h  and 1h fillings, respectively. The corresponding transport spectra are shown in Figs. 4c, 4d, respectively. A perpendicular magnetic field almost restores (for our parameter set) Kramers degeneracy, thus revitalizing the Kondo resonance for this angle. 
As the field is more and more aligned to the CNT's axis, the degeneracy is removed, which also enables us to distinguish between ${\cal P}$ and ${\cal C}$ transitions. As in the axial case of Fig. 2, only the inelastic resonance associated to the ${\cal C}$ transition is clearly resolved in both the experiment and theory. \\

%
%

 {\bf Entropy and specific heat.} Usually, quantum entanglement suffers from decoherence effects \cite{Buchleitner2009,Akulin2005}. The Kondo-Kramers singlets,  however, are associated to a global symmetry of the quantum dot-plus lead complex,  and are robust against thermal fluctuations or finite bias effects as long as the impurity is in the Fermi liquid regime \cite{Hewson1997} ($T< 0.01$ $ T_{\rm K}$ for our experiment).
For larger energy scales, $0.01< T/T_{\rm K} <1$  the impurity is not fully screened, but Kondo correlations persist yielding universal behavior of relevant  observables, as seen e.g. in Fig. 1h at the level of the linear conductance.   In order to further investigate the impact of  thermal fluctuations on Kondo correlations, we have calculated the temperature dependence of the  impurity entropy
$S_{\rm CNT}=S_{\rm tot}-S_{\rm L}$, where the $S_i$ are thermodynamic entropies, and of the impurity specific heat  \cite{Merker2012} (see Supplementary Note 1 and Supplementary Figure 2). The conditional entropy $S_{\rm CNT}(T)$  remains close to zero up to temperatures $T\approx  0.01$ $T_{\rm K}$, indicating that the system is to a good approximation in the singlet ground state. At higher temperatures the impurity entropy grows, but universality is preserved up to temperatures close to $ T_{\rm K}$, at which the entropy approaches the value $k_{\rm B} \log 2$.

\section*{\large D\MakeLowercase{iscussion}} 
Our results  show  that specific low-energy inelastic processes, observed in the perturbative cotunneling regime,  
  tend to be blocked  in the Kondo regime due to antiferromagnetic-like correlations, which  at zero temperature yield a many-body ground state with net zero Kramers pseudospin.
This signature of the Kondo effect  is  universal, in the sense that it does not depend on the degree of the spin-orbit coupling  or valley mixing specific to a given CNT. As such, it is also expected for \SUF\ correlated CNTs, which explains the missing inelastic resonance in the seminal work \cite{nature-jarillo:484}.
Furthermore, we believe that such pseudospin selective suppression  should be detectable also in a variety of other tunable  quantum dot systems with emergent \SUF\ and  \SUT\ $\otimes$ \SUT\ Kondo effects  \cite{Borda2003,Sasaki2004,Tettamanzi2012,Minamitami2012,Keller2013,Crippa2015}. 

 Because the screening is progressively suppressed by increasing the temperature or the bias voltage, it should be possible to recover such inelastic transitions by continuosly tuning those parameters. Indeed, signatures of the re-emergence of the ${\cal P}$ transition are seen in the KEA calculations and experimental traces at fields around 0.9 T in the form of an emerging shoulder, see Supplementary Figure 3. Experiments at larger magnetic fields, not accessible to our experiment,  are required to record the evolution of this shoulder, and thus the suppression of (non-equilibrium) Kondo correlations by an applied bias voltage.

\section*{\large M\MakeLowercase{ethods}}
{\bf Experimental fabrication.}
Devices were fabricated from degenerately doped silicon ${\rm Si/SiO_{2}/Si_{3}N_{4}}$ wafers with a 500 nm thick thermally grown ${\rm SiO_{2}}$ layer and 50 nm ${\rm Si_{3}N_{4}}$ on top. Metal leads separated by 200 nm were first defined by electron-beam lithography and deposited using electron-gun evaporation. A thickness of 2 nm Cr followed by 50 nm Pt was used. A 200 nm deep trench was created using both dry-etching and wet-etching. A second step of electron-beam lithography was used to design a 50 nm thin metallic local gate at the bottom of the trench. Catalyst was then deposited locally on top of the metal leads. Carbon nanotubes were then grown by the CVD technique to produce as clean as possible devices. Only devices with room temperature resistances below 100 k$\Omega$ were selected for further studies at very low temperature.  A scanning electron microscopy of a device similar to the one measured in this work is shown in the Supplementary Figure 1.\\

{\bf Transport methods.} For the transport calculations, three different approaches have been used: the density-matrix numerical renormalization group (DM-NRG) method, a real time diagrammatic perturbation theory (PT) for the dynamics of the reduced density,  and the analytical   Keldysh effective action (KEA) approach. 
Further details are discussed in the Supplementary Note 2.  \\

{\bf Model CNT Hamiltonian.}
In our calculations we have used  the standard model Hamiltonian for the longitudinal mode of a CNT accounting for spin-orbit coupling (SOC), valley mixing,  onsite and exchange Coulomb interactions, and an external magnetic field  \cite{Laird2015}. Regarding both SOC and the valley mixing as perturbations breaking the \SUF\ symmetry of the single particle CNT Hamiltonian,  it has the general form 
\begin{equation}
  \Hop_{\CNT}=\Hop_{\dtxt}+\Hop_{\SO}+\Hop_{\rm KK'}+\Hop_{\rm U}+\Hop_{\rm J}+\Hop_{\B},
\label{eq:H_CNT}
\end{equation}
where $\Hop_{\dtxt}+\Hop_{\rm U}$  is the \SUF\ invariant component. In the basis set  $\{K' \uparrow,K' \downarrow, K \uparrow, K \downarrow\}$ indexed by the valley and spin degrees of freedom $\tau =K',K=\pm$ and $\sigma=\uparrow,\downarrow=\pm$, respectively,  it reads
\begin{equation}
  \Hop_{\dtxt}+\Hop_{\rm U}=\varepsilon_{\dtxt}\sum_{\tau,\sigma=\pm}\dop^{\dagger}_{\tau,\sigma}\dop_{\tau,\sigma}+\frac{U}{2}\sum_{(\tau,\sigma)\neq (\tau',\sigma')}\hat n_{\tau,\sigma}\hat n_{\tau',\sigma'},
\end{equation}
with $\varepsilon_{\dtxt}$ the energy of the quantized longitudinal mode, which can be tuned through the applied gate voltage, and $U$ accounting for charging effects.  
Valley mixing and SOC break  the \SUF\ symmetry with characteristic energies \DKK\ and \DSO, respectively. The corresponding contributions read: 
\begin{equation}
  \label{eq:dot_hamiltonian}
   \Hop_{\rm KK'}+ \Hop_{\SO}=\frac{\Delta_{\KK}}{2}\sum_{\tau,\sigma=\pm}\dop^{\dagger}_{\tau,\sigma}\dop_{-\tau,\sigma} + 
  \frac{\Delta_{\SO}}{2}\sum_{\tau,\sigma=\pm}\sigma\tau\hat n_{\tau,\sigma}.
\end{equation}
The SOC term is a result of the atomic spin-orbit interaction in carbon, and thus exists also for ideally infinitely long CNTs \cite{Ando2000}. The valley mixing, in contrast, is absent in long and defect free CNTs. It only arises due to scattering off the boundaries in finite length CNTs or due to disorder \cite{Kuemmeth2008,Izumida2015,Marganska2015}.  
It is expected to be zero in disorder-free CNTs of the zig-zag class, due to angular momentum conservation rules, and finite in CNTs of the armchair class \cite{Marganska2015}. In our experiments, according to Table \ref{tab:parameters}, the valley mixing is very small, which suggests a tube  of the zig-zag class.   

Similar to the SOC and valley mixing, the exchange interaction preserves time reversal symmetry.
Its microscopic form is not known for abritrary chiral angles. It has been evaluated so far for the case of pure armchair tubes \cite{Mayrhofer2008}, and for the zig-zag class  \cite{Secchi2009,Laird2015} CNTs. Because the experiments suggest that our tube is of the zig-zag class, we choose in the following a form suitable to describe this case.
It reads  
\begin{equation}
   \Hop_{\rm J}=-\frac{ J}{2}\sum_{\sigma=\pm}\{\hat n_{K,\sigma}\hat n_{K',\sigma}+ \dop^{\dagger}_{K,\sigma} \dop^{\dagger}_{K',-\sigma}\dop_{K,-\sigma}\dop_{K',\sigma}\},
\end{equation}
with $J<0$ the exchange coupling.
%
%
%
%
Finally, contributions arising from a magnetic field {\bf B} contain both Zeeman and orbital parts. Decomposing {\bf B} into components parallel and perpendicular to the tube axis,
$B_\parallel=B\cos\theta$  and $B_\perp=B\sin\theta$, respectively, one finds:
\begin{eqnarray}
    \Hop_{\B}&=&\Hop_{\B }^{\rm Z}+\Hop_{\B}^{\rm orb}\nonumber\\
    &=&B_\parallel\sum_{\tau,\sigma=\pm}\big(\frac{g_{\rm s}}{2}\mu_{\rm \B}\sigma 
+\mu_{\rm orb}\tau \big)\dop^{\dagger}_{\tau,\sigma}\dop_{\tau,\sigma}\nonumber\\
&+&\frac{g_{\rm s}}{2}\mu_{\rm B}B_{\perp}\sum_{\tau,\sigma=\pm}\dop^{\dagger}_{\tau,\sigma}\dop_{{\tau},-\sigma}.
\end{eqnarray}
Notice that the spin and valley remain  good quantum numbers in the presence of an axial field ($\theta=0,\pi$),  while a perpendicular component flips   the spin degrees of freedom. 
The parameters of the CNT Hamiltonian used to fit the experimental data shown in Figs. 1, 2 and 4  are listed in Table \ref{tab:parameters}. \\

{\bf  Kramers charge and pseudospin representation. }
We call Kramers basis the quadruplet $\{\vert i\rangle\}$,  $ i=1,2,3,4$  (shown in Fig. 1b ) which diagonalizes the single particle part 
 $\hat H_{0}=\Hop_{\rm d}+\Hop_{\KK} +\Hop_{\SO} + \Hop_{ \B}$  of the CNT Hamiltonian.
For magnetic fields parallel or perpendicular to the CNT axis, this Hamiltonian is easily diagonalized, see e.g. \cite{Schmid2015}. For other orientations of the field, because of the combined action of SOC and  valley mixing, such states are a linear superposition of all the basis states $\{\vert \tau,\sigma \rangle \}$, such that neither the spin nor the valley are in general good quantum numbers any more.  
One has to resort to numerical tools to find both the eigenvectors $\{ \vert i \rangle \}$ and the eigenvalues $\varepsilon_i$, $i=1,2,3,4$. 
The angular dependence of these eigenenergies is sketched in Fig. 4. 

Despite the complexity inherent in the Hamiltonian $\hat H_{0}$, a closer inspection reveals the existence of conjugation relations among the quadruplet of states $i=1,2,3,4$ 
generated by the time-reversal operator  $\hat{\cal T}$, as well as by the particle-hole like and chirality operators $\hat{\cal P}$ and $\hat{\cal C}=\hat{\cal P}\hat{\cal T}^{-1}$, respectively \cite{Schmid2015}. Specifically, the states are ordered such that  $(1,2)$ and $(3,4)$ are time-reversal partners, while $(1,4)$ and $(2,3)$ are particle-hole partners.
In the  $\{ \vert \tau,\sigma \rangle \}$ basis the operators read 
\begin{eqnarray}
\label{eq:CPT}
    \hat{\cal T}&=&\hat\kappa\sum_{\tau,\sigma}\sigma\dop^{\dagger}_{-\tau,-\sigma}\dop_{\tau,\sigma},\\
    \hat{\cal P}&=&\hat\kappa\sum_{\tau,\sigma}\sigma\tau\dop^{\dagger}_{-\tau,\sigma}\dop_{\tau,\sigma},\\
    \hat{\cal C}&=&\sum_{\tau,\sigma} (-\tau)\dop^{\dagger}_{\tau,-\sigma}\dop_{\tau,\sigma},
\end{eqnarray}
where 
$\hat\kappa$ stands for the complex conjugation operator.  
 In the absence of a magnetic field $\hat{\cal T}$ commutes with the total CNT Hamiltonian, yielding a single-particle spectrum with two  degenerate Kramers doublets (1,2) and (3,4) separated by the inter-Kramers splitting $\Delta=\sqrt{\Delta_{\rm SO}^2+\Delta_{\rm KK'}^2}$ (see Fig. 1b). 
As far as the $\hat{\cal P}$ and $\hat{\cal C}$ operators are concerned,  at zero magnetic field they are symmetries only in the absence of SOC and valley mixing. Since both anticommute with  $\hat H_{\rm SO}+\hat H_{\rm K K'}$, it holds for ${\cal P}$-conjugated pairs,  $\varepsilon_{1,2}(\Delta)=\varepsilon_{4,3}  (-\Delta)$.
 A magnetic field breaks the time-reversal symmetry; however, because  $\hat H_{ \B}$ anticommutes with $\hat {\cal T}$, formerly degenerate Kramers states are still related to each other by Kramers conjugation. 
 For an arbitrary magnetic field {\bf B} time-reversal conjugation and particel-hole conjugation imply \cite{Schmid2015}:
\begin{eqnarray}
\varepsilon_{1,4}({\bf B})&=&\varepsilon({\bf B}) \pm \frac{1}{2} \Delta ({\bf B}),\\
\varepsilon_{2,3}({-\bf B})&=&\varepsilon_{1,4}(\bf{B}),
\end{eqnarray}
where $\varepsilon ({\bf B})$ and $\Delta ({\bf B})$ reduce to the longitudinal energy and Kramers splitting $\varepsilon_{\rm d}$ and $\Delta $, respectively,  at zero field.

These relations clearly suggest the introduction of auxiliary charge  $\hat N_{ij}:=\hat n_i+\hat n_j$ and pseudospin $\hat J^z_{ij}=(\hat n_i-\hat n_j)/2$ operators, in terms of which we can write
\begin{eqnarray*}
\hat H_{0} &=&\varepsilon({\bf B})\hat N_{14}+\Delta({\bf B})\hat J_{14}^z +\varepsilon(-{\bf B}) \hat N_{23} +\Delta (-{\bf B})\hat J^z_{23} . 
\end{eqnarray*}
Introducing the average quantities $\bar\Delta ({\bf B}) :=(\Delta ({\bf B})  + \Delta (-{\bf B}) )/2$, $\bar\varepsilon ({\bf B}):= (\varepsilon ({\bf B})  + \varepsilon (-{\bf B}) )/2$, as well as the differences $\delta\Delta ({\bf B}) :=(\Delta ({\bf B})  - \Delta (-{\bf B}) )/2$, $\delta\varepsilon ({\bf B}):= (\varepsilon ({\bf B})  - \varepsilon (-{\bf B}) )/2$,
the CNT Hamiltonian can  be easily recast in terms of total charge and pseudospin  of a Kramers pair. It reads:
\begin{eqnarray*}
\hat H_{0}&=& \left(\bar\varepsilon({\bf B}) +\frac{\bar\Delta ({\bf B})}{2}\right) \hat N_{12}+[2\delta\varepsilon ({\bf B}) +\delta\Delta({\bf B})]\hat J_{12}^z\\ 
&+& \left(\bar\varepsilon({\bf B}) -\frac{\bar\Delta ({\bf B})}{2}\right) \hat N_{43}+[2\delta\varepsilon ({\bf B}) -\delta\Delta({\bf B})]\hat J_{43}^z .
\end{eqnarray*}
Such equation is   Eq. (1) in the main part of the manuscript upon calling $\hat J^z_{43}=\hat J^z_{\rm d}$, $\hat J^z_{12}=\hat J^z_{\rm u}$, and similarly $\hat N_{43}=\hat N_{\rm d}$, 
$\hat N_{12}=\hat N_{\rm u}$.\\

{\noindent \bf Data availability} The data that support the main findings of this study are available  from the corresponding author upon request.\\
%




\section*{\large A\MakeLowercase{cknowledgments}}
The authors acknowledge fruitful discussions with C. Strunk, A. H\" uttel,   G. Zar\'and and C. P. Moca as well as  financial support by the Deutsche
Forschungsgemeinschaft via  SFB~689  
and GRK~1570, and by the ERC Advanced Grant MolNanoSpin No. 226558.

\section*{\large A\MakeLowercase{uthor contributions}}
M.N. performed the perturbative non-equilibrium calculations, S.S. evaluted the differential conductance in the Kondo regime using the non-equilibrium  Keldysh effective action approach, while
 D.M. did the equilibrium DM-NRG simulations. M.M. evaluated the magnetospectrum of the isolated nanotube and devised all the figures. N.-V.N. helped to fabricate and characterize the devices,  J.-P.C. carried out and analyzed the experiments while W.W. supervised them. M.G. performed the theoretical analysis and wrote the manuscript with critical comments provided by all authors.

\section*{\large A\MakeLowercase{dditional information}}
{\noindent \bf Supplementary information} is available in the online version of this paper. \\

{\noindent \bf Competing financial interests}
The authors declare no competing financial interests.\\

\widetext
\newpage

\begin{figure*}
\includegraphics[width=16cm]{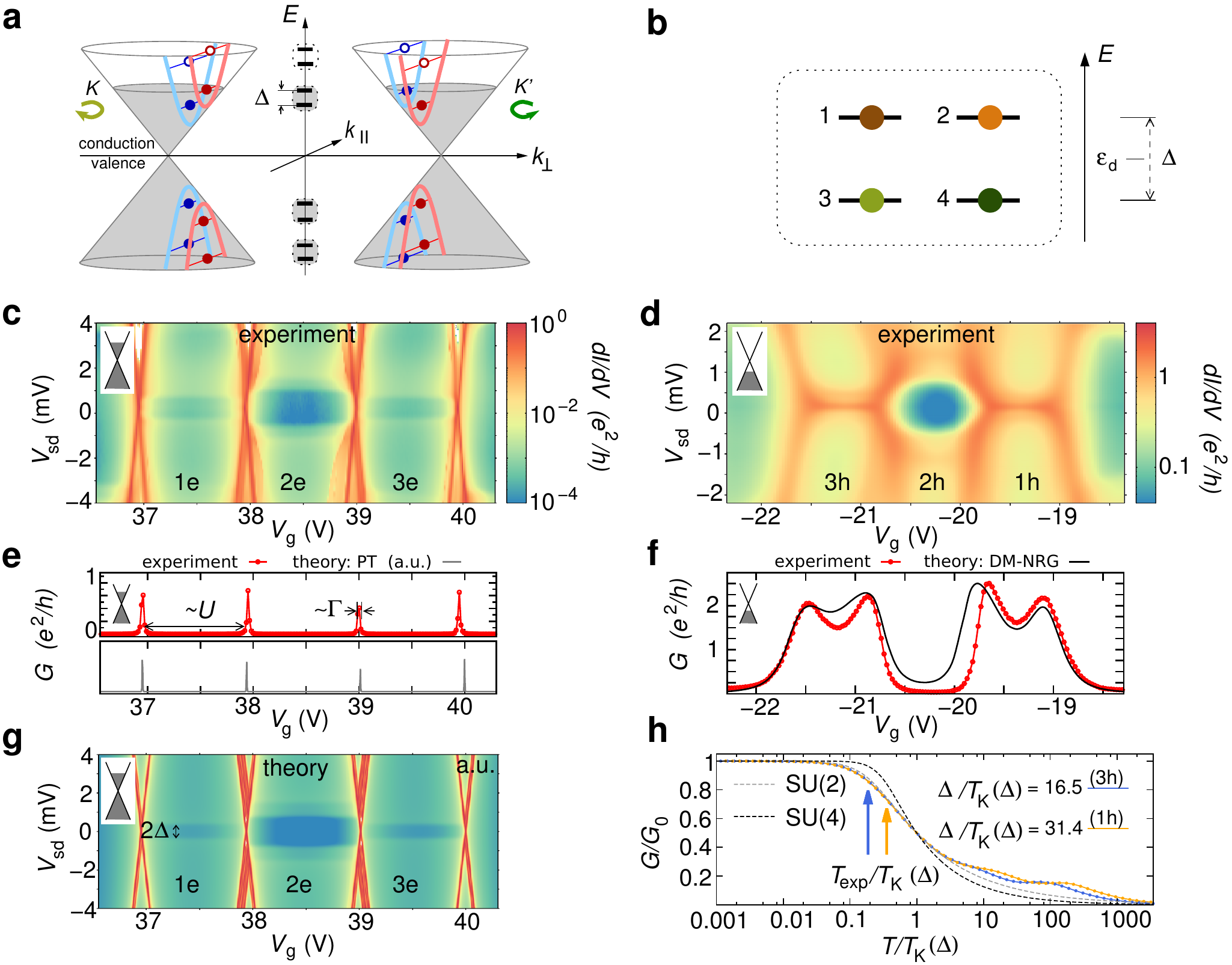}
\caption{\label{fig:ovbias}
\textbf{Transport regimes  and bound states of a CNT quantum dot.} \textbf{a,}  A CNT with spin-orbit coupling is characterized by spin (blue, red) and valley (K,K') resolved transverse modes (blue and red hyberbolae). The CNT chemical potential (upper limit of the shaded regions of the Dirac cones) is adjusted by sweeping the gate voltage from positive values (electron regime) to negative values (hole regime). Quantum confinement yields the quantization  of the longitudinal momentum $k_\parallel$ (empty/solid bullets denote empty/filled bound states).   \textbf{b,} A generic quadruplet of bound states is composed of two Kramers doublets separated by the inter-Kramers splitting $\Delta$. \textbf{c} and \textbf{d,} Experimental stability diagrams demonstrating the successive filling of  a quadruplet with electrons (panel \textbf{c}), and holes (panel \textbf{d}). On the electron side, sequential transport is exponentially suppressed inside the Coulomb valleys; the dominant mechanism is cotunneling (panel \textbf{c}).  The appearance of high conductance ridges at zero bias (panel \textbf{d}) in valleys with odd holes is a signature of the Kondo effect. \textbf{e} and \textbf{f,} Experimental gate traces at zero bias are compared with theoretical predictions obtained with perturbative (panel \textbf{e}) and non-perturbative DM-NRG (panel \textbf{f}) approaches. \textbf{g,} Theoretical stability diagram for the electron side reproducing  the experiment of panel \textbf{c}. \textbf{h,} Scaling behaviour of the linear conductance in the middle of the valleys with odd hole numbers, {$G_0 \approx 2e^2/h$}. The system lies in the crossover regime ($0.1 < T_{\rm exp}/T_{\rm K}(\Delta) <1)$, as pointed out by the arrows. $T_{\rm K}$ is the Kondo temperature determined from the DM-NRG calculation according to $G(T_{\rm K})=G_0/2$.  }
\end{figure*}

\begin{figure*}
\includegraphics[width=16cm]{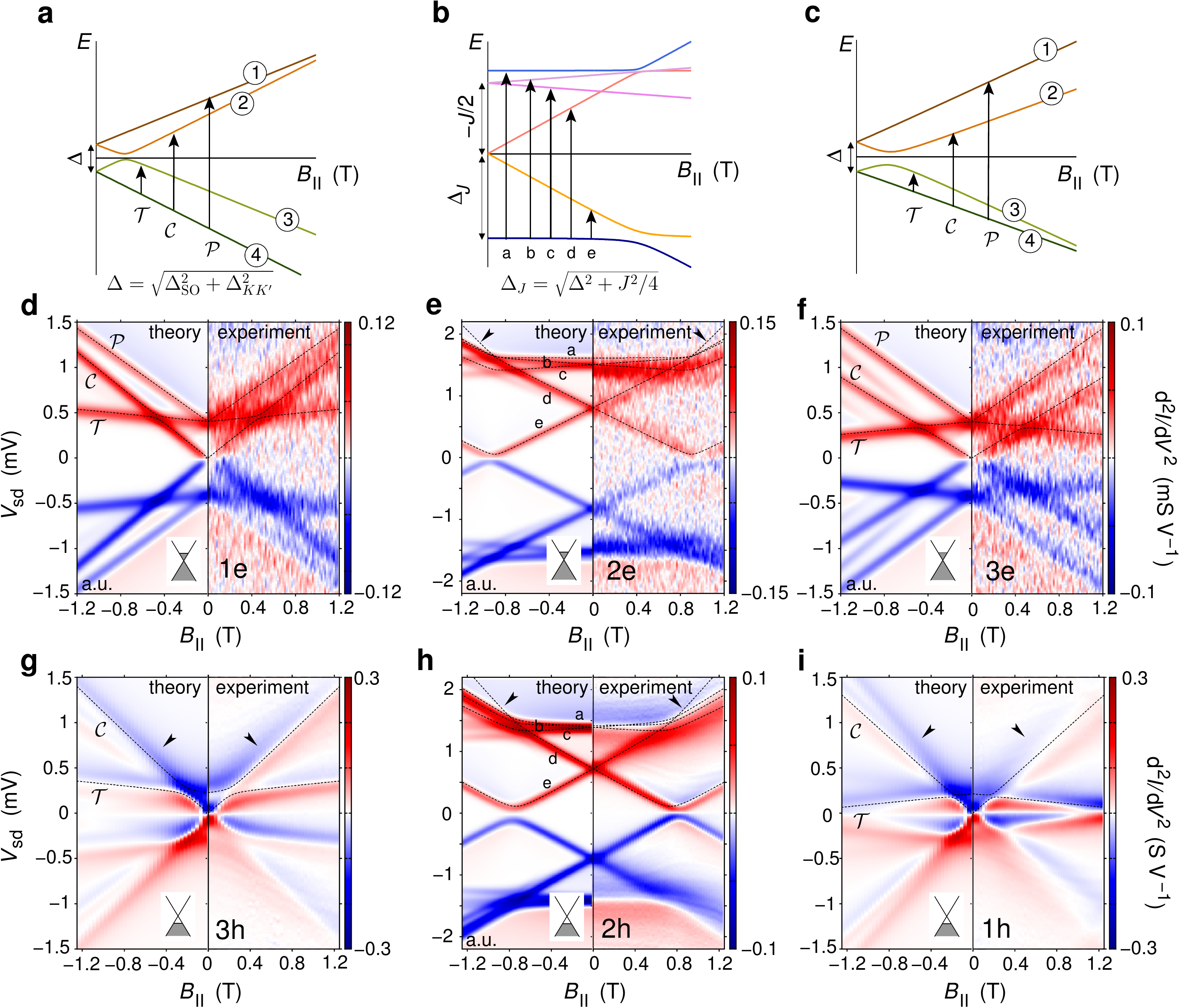}
\caption{\label{fig:tempdep}
\textbf{Energy spectra and  magnetotransport in both cotunneling and Kondo regimes.} \textbf{a-c,} Excitation spectra for  electron filling  ($n_e=1 ,2,3$ from left to right). 
The parameters, \DSO, \DKK\ and $J$  account for SOC, valley mixing and exchange splitting, respectively. \textbf{d-f,}  Current second derivative d$^2I$/dV$^2$ in the electron regime at gate voltages fixed in the middle of the 1e, 2e and 3e charge states, as a function of bias voltage and parallel magnetic field. 
Each panel reports experimental data (positive magnetic field) and transport calculations (negative field). The dotted lines correspond to the transition energies from the ground state calculated directly from the spectra \textbf{a-c}. 
At odd filling (panels \textbf{a} and \textbf{c}), all possible ground state transitions, denoted by $\cal{C}$, $\cal{P}$ and $\cal{T}$ are observed in the experiment (panels \textbf{d} and \textbf{f}). Being signalled by cotunneling steps in the current first derivative, they yield maxima/minima in the second derivative. Likewise for even occupation, except for the ``a'' transition at high field (marked by arrows), forbidden by  selection rules. \textbf{g-i,} d$^2I$/dV$^2$ maps  in the hole regime for the 1h, 2h and 3h charge states. While the experimental results for the 2h and 2e cases are similar, the $\cal{P}$ transitions are no longer experimentally resolved, as predicted by the transport theory, due to the Kondo effect (panels \textbf{g} and \textbf{i}). These missing resonances are indicated by arrows in panels \textbf{g} and \textbf{i}. In the Kondo-regime $\cal {T} $ and $\cal{C}$ transitions yield maxima in the differential conductance, and hence zeroes in the second derivative. Near maxima (minima) of d$I$/dV the second derivative decreases (increases), i.e., it changes from red  to blue (blue to red) upon increasing the bias.    The experimental part of panels \textbf{g}, \textbf{h} and \textbf{i} has  been adapted from \cite{Cleuziou2013}. 
}
\end{figure*}

\begin{figure*}
\includegraphics[width=\textwidth]{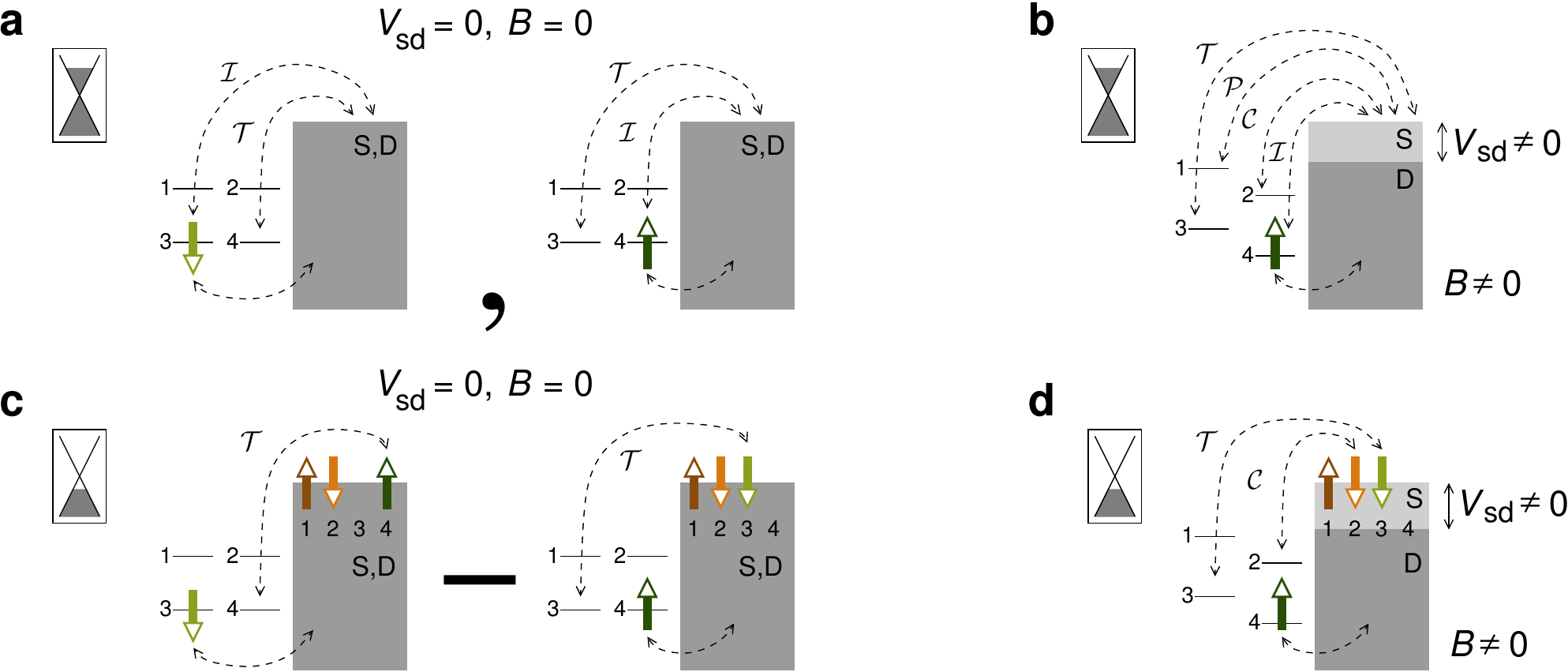}
\caption{\label{fig:ovbias}
\textbf{Ground state configurations and virtual processes  of a CNT quantum dot with one electron filling  in the cotunneling and Kondo regimes}. \textbf{a,} In the cotunneling regime the one-electron ground state is doubly degenerate, with opposite values of the Kramers pseudospin. Elastic cotunneling processes  to source (S) and drain (D) leads (grey areas) involving the same pseudospin, ${\cal I}$, and its Kramers partner, ${\cal T}$, contribute to the linear transport. \textbf{b,} Kramers degeneracy is broken by a  magnetic field. A finite bias allows us to identify the three inelastic processes $\cal{T}$, $\cal{P}$ and $\cal{C}$ which connect the bound states within a quadruplet. \textbf{c,} The ground state in the Kondo regime is a singlet with no net Kramers pseudospin. Only virtual ${\cal T}$ fluctuations which involve a pseudospin-flip matter at low energies. \textbf{d,} At finite bias voltages the inelastic $\cal{T}$, $\cal{C}$ transitions, which involve a pseudospin flip, are the most relevant in the deep Kondo regime.}
\end{figure*}

\begin{figure*}
\includegraphics[width=16cm]{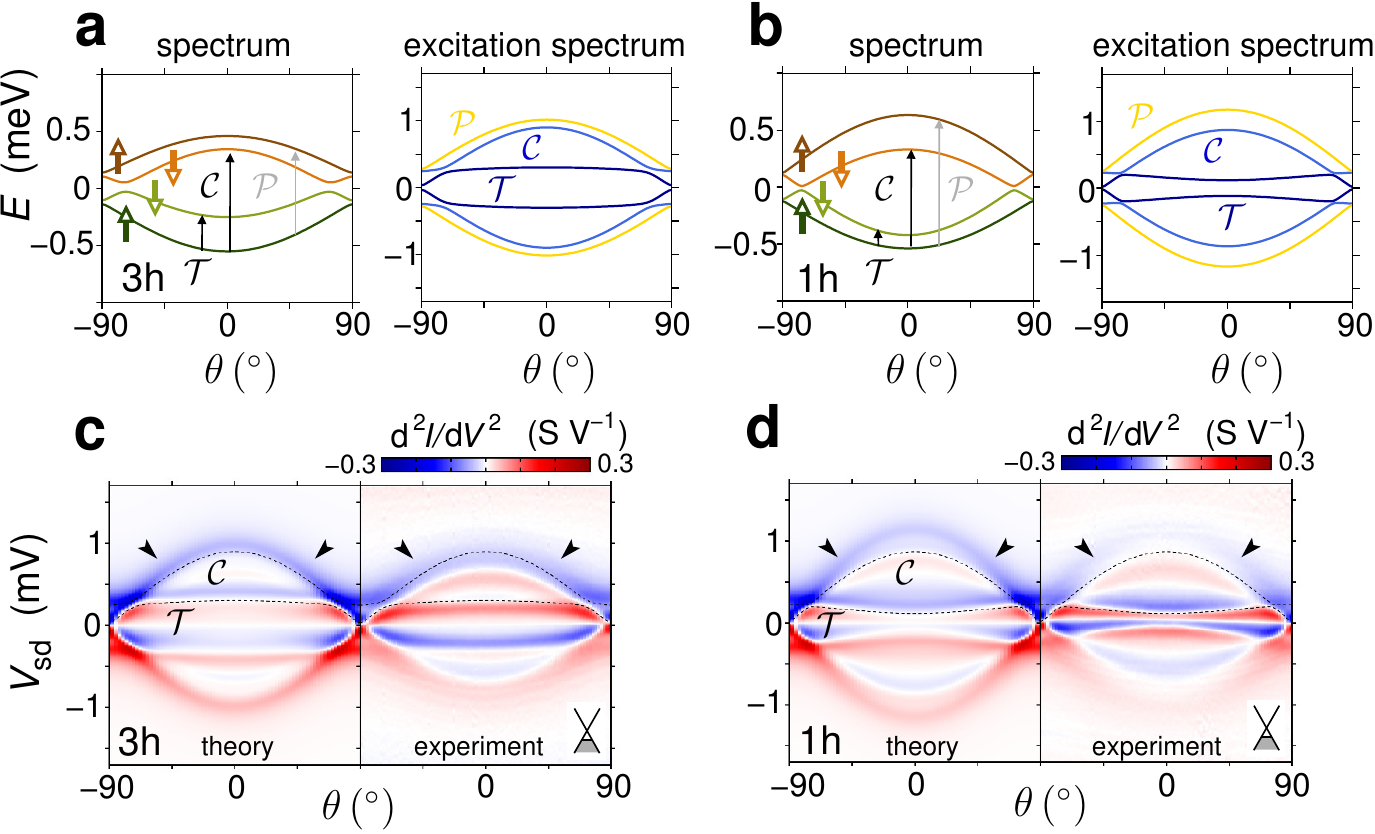}
\caption{\label{fig:magnetic}
\textbf{Angular dependence of both spectrum and transport characteristics as the magnetic field is rotated in the CNT plane. a,b,} Sketch of the spectrum and excitation spectrum at 3h and 1h fillings, respectively, as a function of the polar angle $\theta$ formed by an applied magnetic field and the CNT axis. A classification of the inelastic transtions according to the $\cal{T}$, ${\cal C}$ and ${\cal P}$ operations is still possible. {\textbf{c,d,}} As-measured  and KEA transport calculations for the current's second derivative d$^2I$/dV$^2$.  The absence of ${\cal P}$ transitions is independent of the direction of the applied field. The experimental part of panel  {\bf d} has been adapted from \cite{Cleuziou2013}. The magnetic field magnitude in panels  {\bf a-d} is set to 0.8 T.}
\end{figure*}

\clearpage
\newpage
\begin{table}[t]
 \begin{tabular}{l c c c c}
 \hline
& & & holes   & electrons  \\
& & &  (shell $N_h$=6) &  (shell $N_e$=6) \\
   \hline
   \DSO (meV) & & & -0.21 & -0.4 \\
   \DKK (meV) & & & 0.08 & 0.04 \\ 
  \muorb (meV/T) & & & 0.51 (3h), 0.51 (2h), 0.55 (1h) &  0.43  \\
\hline
    $U$ (meV) PT & & &  &  26,5  \\ 
  $U$ (meV) NRG & & &  4,7 &  \\ 
  $U$ (meV) KEA & & & $\infty$ (3h, 1h) &  \\ 
    $J$ (meV) PT & & & -1.35 &  -1.4  \\
   \hline 
$\Delta_\mu B_\parallel$ (meV/T) & & & -0.05 & -0.06 \\
   $e\Delta V_{\rm sd}$ (meV) & & & 0.12 & 0.28 \\
   \hline
   \end{tabular}
 \caption{\label{tab:parameters} {\bf Parameter set}. The table shows the parameters used to fit the electronic transport spectra of the CNT  in the gate voltage region shown in the main text. It corresponds to the  valence quadruplet $N_h$=6 (hole transport),  and the  conduction quadruplet $N_e$ =6 (electron transport), counting the Coulomb diamonds from the band gap.  The abbreviations PT, NRG and KEA refer to the three theoretical methods used in our calculations (see text). The experimental data for each Coulomb valley are offset by $\Delta V_{\rm sd}$, and tilted in the magnetic field by $\Delta_\mu B_\parallel$, resulting in an asymmetry between the measurement in fields parallel and antiparallel to the CNT axis. In all the plots presented in the work both the offset and the tilt have been removed. }
 \end{table}

\end{document}